# Comment on "Left-Handed Materials Do Not Make a Perfect Lens"

JB Pendry
*The Blackett Laboratory, Imperial College, London, SW7 2BU, UK.*

In a recent Physical Review Letter [1] Garcia and Nieto Vesperinas (GNV) dispute the claim of perfect lensing made in [2]. The thrust of the GVN paper is that the solutions proposed in [2] imply infinite energy density and are therefore inadmissible. They claim that finite absorption leads to catastrophic collapse of the amplifying solutions vital to focussing and that no useful effect can be achieved.

In this Comment I show that, on the contrary, careful consideration of absorption results in solutions that are always well behaved and evolve smoothly and continuously to perfect resolution in the limiting case of zero absorption. Contrary to assertions in [1] the original Letter [2] took losses fully into consideration; one of several misattributions in [1].

First consider the semi-infinite case shown in Figure 1. An object outside a negative medium resonantly excites surface plasmons. We allow for the negative index medium (NIM) to be slightly lossy.

$$\varepsilon = -1 + i\delta \quad \mu = -1 + i\delta \tag{1}$$

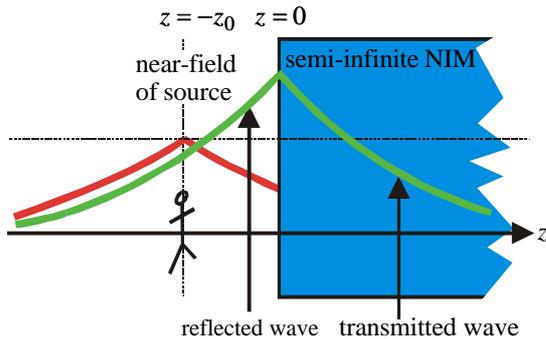

Figure 1. An source of electromagnetic fields excite surface plasmons at the vacuum/NIM interface.

Some of the field is transmitted into the slab and some of the field is reflected, as calculated in [2]. For any given value of the parallel wave vector ($k_y$),

$$t \approx \frac{-2}{i\delta}, \quad r \approx \frac{-2}{i\delta}, \quad \delta \ll 1 \tag{2}$$

Note that all the fields decay away exponentially to infinity and are normalisable in real space. The discontinuity in source fields is accounted for by charges and currents within the source plane. Note that the fields proposed in reference [2] are not correctly described in [1]. Furthermore when we sum over all wavevectors to give the total wavefield, that is also normalisable for any physical source, as can easily be demonstrated from (2).

Next consider a finite slab where there are two plasmon wavefields to be excited which interact one with the other (figure 2). The field multiply scatters between the two plasmons and it is this effect which is at the heart of the perfect lens as it suppresses the first surface plasmon.

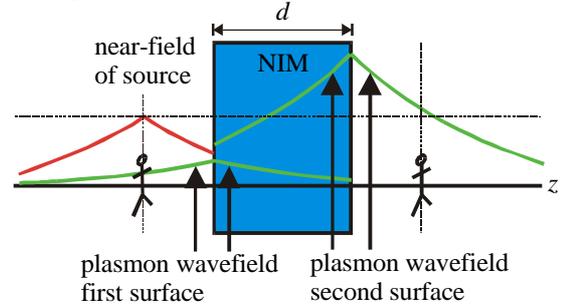

Figure 2. As for figure 1 but with a finite slab of NIM. In the lossless limit of $\delta \to 0$ the first surface plasmon has zero amplitude and only the second surface is excited, refocusing the image at a distance $2d$ from the source.

In the lossless limit of $\delta \to 0$ the second surface plasmon dominates, the slab amplifies the wave field for each $k_y$, and the result is a focussed image. However this process is reversed if,

$$\delta > \exp\left(-\sqrt{k_y^2 - \omega^2 c_0^{-2}}\, d\right) \tag{3}$$

when there is an exchange of roles between the two surface plasmons: if (3) holds the first surface plasmon dominates and amplification is replaced by attenuation. Condition (3) ensures that once again that the total wavefield is normalisable for any physical source. The transition between the two regimes is a smooth one and if $\delta \ll 1$ many wave vectors contribute to the image enabling sub-wavelength imaging with spatial resolution,

$$\Delta \approx (-\ln \delta)^{-1} \tag{4}$$

Therefore my conclusions are contrary to those of GNV: finite absorption prevents the divergences to which GNV object and although absorption limits the ultimate resolution of the lens, in any event there are no theoretical to obtaining resolution far beyond the conventional limitations of wavelength. The real challenge is the practical one of designing very low loss NIM's.